\documentclass[preprint,12pt]{elsarticle}

\usepackage{graphics}

\usepackage{amssymb}

\begin{document}

\begin{frontmatter}



\title{Asset returns and volatility clustering in financial time series}


\author[as]{Jie-Jun Tseng}
\ead{gen@phys.sinica.edu.tw}
\author[as,ut]{Sai-Ping Li}

\address[as]{Institute of Physics, Academia Sinica, Nankang, Taipei 115, Taiwan}
\address[ut]{Department of Physics, University of Toronto, Toronto, Ontario M5S 1A7, Canada}

\begin{abstract}
An analysis of the stylized facts in financial time series is carried out.
We find that, instead of the heavy tails in asset return distributions,
the slow decay behaviour in autocorrelation functions of absolute returns
is actually directly related to the degree of clustering of large fluctuations
within the financial time series.
We also introduce an index to quantitatively 
measure the clustering behaviour of fluctuations in these time series and show that
big losses in financial markets usually lump more severely than big gains.
We further give examples to demonstrate that comparing to conventional methods,
our index enables one to extract more information from the financial time series. 
\end{abstract}

\end{frontmatter}

%
\section{Introduction}
\label{sec:1}
In financial markets, prices of stocks and commodities fluctuate over time 
which then produce financial time series.
These time series are in fact of great interest both to practitioners and theoreticians
for making inferences and predictions.
Using modern day technologies, one can now obtain a vast amount of financial data
that record every transaction in financial markets which was not possible a couple of decades ago.
The analysis involved is also far more complicated.
With the tremendous amount of information obtained over the past decade,
researchers have now come to agree on several stylized facts about financial markets,
i.e., heavy tails (or fat tails in the terminology of finance) in asset return distributions,
absence of autocorrelations of asset returns,
volatility clustering, aggregational normality
and asymmetry between rises and falls~\cite{ref:sf1,ref:sf2,ref:sf3,ref:EPBOOK1,ref:EPBOOK2}. 
\begin{figure}
  \centering{\includegraphics[width=0.8\textwidth]{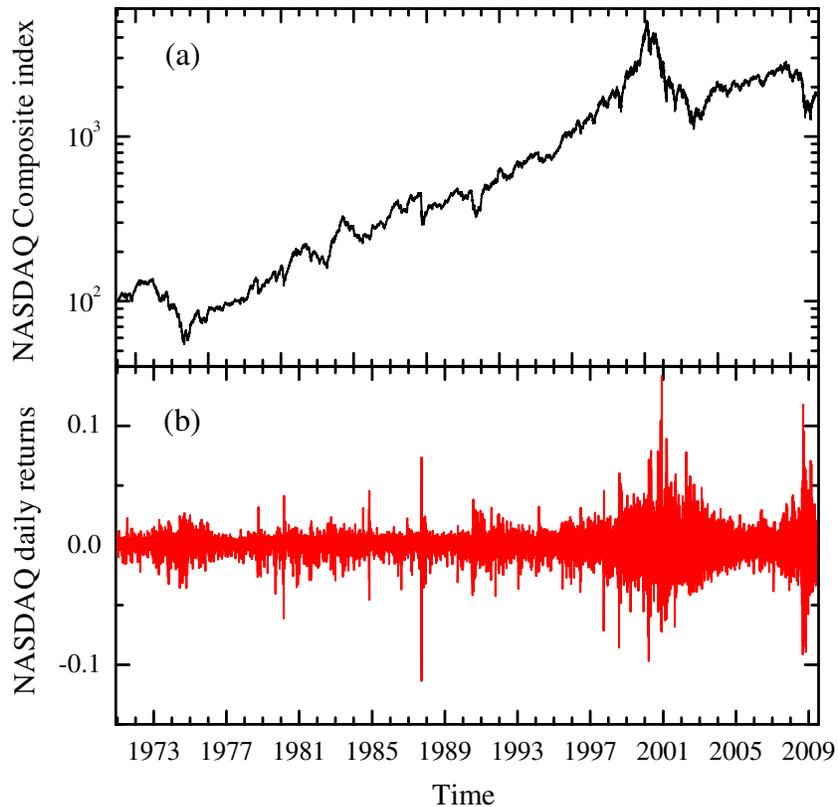}}
  \caption{\label{fig:nasdaq:return}
  The empirical data of the NASDAQ Composite index from February 8, 1971
  through June 30, 2009. (a) shows the historical daily closing price
  while (b) plots the daily returns during this period.}
\end{figure}
Figure~\ref{fig:nasdaq:return} (a) shows a plot of the historical daily closing values of
NASDAQ Composite index from February 8, 1971 through June 30, 2009
while figure~\ref{fig:nasdaq:return} (b) is its daily price returns during this period.  
The price return $R_\tau(t)$ at time $t$ is defined as the difference
between the price $p(t)$ of a financial asset
(here it is the index value of NASDAQ) at time $t$
and its price a time $\tau$ before, $p(t-\tau)$, divided by $p(t-\tau)$, 
\begin{equation}\label{eq:return}
  R_\tau(t) = \frac{p(t) - p(t-\tau)}{p(t-\tau)} \,\, .
\end{equation}
Therefore, one can obtain the daily returns $R_{1}(t)$ by setting $\tau=1$ trading day and
these returns reflect the price fluctuations in this time series.
We will use daily returns to define fluctuations in a financial price series throughout this article.
As one can see in figure~\ref{fig:nasdaq:return} (b) that the daily returns are varying over time.
A naive thinking would be that these fluctuations are independent, identically distributed (iid)
variables generated by some random processes (i.e., random walks~\cite{ref:bachelier})
and therefore the probability density function of the returns should follow a Gaussian distribution.  
However, it turns out that the empirical distributions of the returns are indeed heavy-tailed.
\begin{figure}
  \centering{\includegraphics[width=0.8\textwidth]{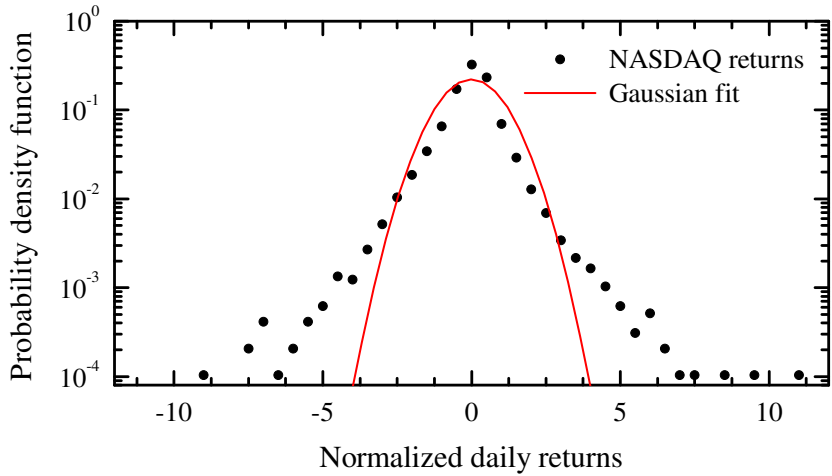}}
  \caption{\label{fig:nasdaq:pdf}
  The probability density function of the normalized daily returns of the NASDAQ 
  index in figure~\ref{fig:nasdaq:return}.}
\end{figure}
In figure~\ref{fig:nasdaq:pdf},
we depict the probability density function of normalized daily returns of the NASDAQ index.
The normalized daily return is defined as $\left(R_1(t)-\mu_{R} \right)/\sigma_{R}$,
where $\mu_{R}$ and $\sigma_{R}$ denote the average and the standard deviation of $R_1(t)$.
One can clearly see that there are heavy tails at the two ends of the distribution.
For comparison, we also include a Gaussian fit with $\mu=0$ and $\sigma=1$.
This is one of the stylized facts that was discovered back in 1960s~\cite{ref:mandelbrot,ref:fama}.
Many studies have been carried out over the years on different financial time series
and the heavy tails in return distributions have always been observed.
There have been many suggestions on the form of the distributions
but no general consensus has been reached on the exact form of the tails so far.
We will not continue our discussion on this issue here
but refer our reader to the literature~\cite{ref:EPBOOK1,ref:EPBOOK2,ref:pdf1,ref:pdf2}.  

\begin{figure}
  \centering{\includegraphics[width=0.8\textwidth]{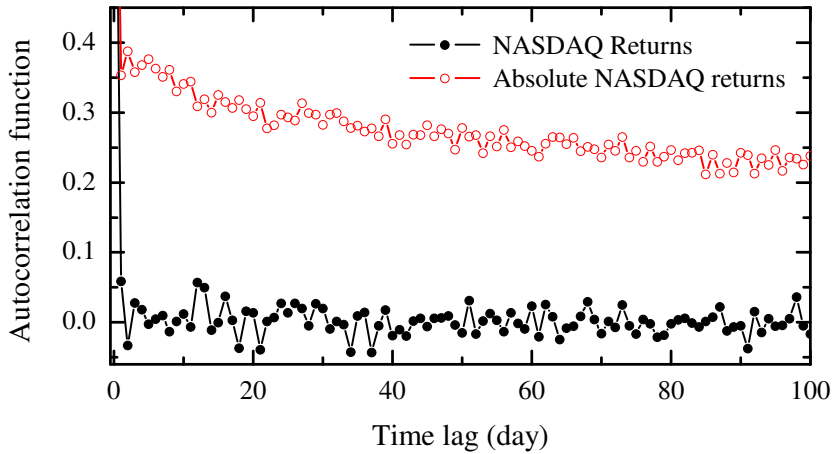}}
  \caption{\label{fig:nasdaq:af}
  The autocorrelation functions of the returns and its absolute value.}
\end{figure}
In addition to those heavy tails in return distributions,
large fluctuations in prices seem to lump together as well~\cite{ref:vc1,ref:vc2}.
If one examines the empirical time series shown in figure~\ref{fig:nasdaq:return},
it is easy to observe that large fluctuations in prices are more often 
followed by large ones while small fluctuations are more likely 
followed by small ones.
This stylized fact is known as volatility clustering~\cite{ref:vc3}.
In financial time series, it is not just that there are more large fluctuations
than pure random processes but also these large fluctuations tend to cluster together.  
It is often suggested that a more quantitative way to view this property
is to look at the autocorrelations of the return series~\cite{ref:vc2}.
The autocorrelation function $C\left(x_t,x_{t+\tau}\right)$ is defined as
\begin{eqnarray}
  C\left(x_t,x_{t+\tau}\right) \equiv
    \frac{\left<\left(x_{t}-\left<x_{t}\right>\right) \left(x_{t+\tau}-\left<x_{t+\tau}\right>\right) \right>}
    {\sqrt{\left<{x_t}^2\right> - \left<x_t\right>^2} \sqrt{\left<{x^2_{t+\tau}}\right> - \left<x_{t+\tau}\right>^2}} \,\, ,
\end{eqnarray}
where $\left<x\right>$ denotes the expectation value of the variable $x$.
While the returns themselves do not show the evidence of temporal correlations,
the absolute returns or their squares do display a positive,
pronounced slowly decaying autocorrelation which indeed exhibit power-law decay behaviour.
The autocorrelations of the absolute value or the square,
etc of the asset returns are often known as the nonlinear autocorrelations.
We will only consider the autocorrelation of the absolute returns
as an example of the nonlinear autocorrelation in this paper.  

Figure~\ref{fig:nasdaq:af} are plots of the autocorrelation functions
of the returns and its absolute value for the time series shown in 
figure~\ref{fig:nasdaq:return}.
It is easy to see that there is no correlation
among the returns since the autocorrelation function drops
to the noise level within a couple of days.
On the other hand, the autocorrelation function of the absolute returns, 
i.e., the nonlinear autocorrelation does exhibit a much slower decay behaviour.
Researchers have fitted this with a power-law decay,
and it is not clear at this moment whether the slow decay should imply long time memory
of the financial time series~\cite{ref:vc3}.
However, one should also keep in mind that
if the time series do possess the properties of the long time memory
and the heavy-tailed distribution, many standard estimation procedures
(i.e., examining sample autocorrelations.) may fail to work~\cite{ref:vc3,ref:vc4,ref:vc5}.
Therefore, in order to have a more reliable measurement
of the volatility clustering, an alternative approach is also needed
while dealing with financial time series.
For instance, if only the clustering behaviour is concerned,
one can simply characterize this property by the concept of probability.
Table~\ref{tab:probability} is an example 
which shows the probability of the occurrence of large and small
fluctuations following the occurrence of large or small fluctuations
on the previous day.
By large (small) fluctuations, we here choose them to be the
largest (smallest) 20\% of all the returns
and the remaining returns are denoted as the rest.
Therefore, each row in table~\ref{tab:probability} sums to unity.
It is easy to see that the probability that
there will be a large (small) return following
a large (small) one on the previous day is significantly higher 
(larger than 20\% in this case) than that of a pure random process.
\begin{table}
  \caption{\label{tab:probability}
  The probability of the occurrence of large and small fluctuations following
  the occurrence of large or small ones on the previous day (the first column).
  The result here is for NASDAQ time series.}
  \begin{center}\begin{tabular}{@{}ll|lll}
      \hline
      &20\% &Largest& Smallest& Rest\\
      \hline
      &Largest & 0.3947& 0.1156& 0.4897\\
      &Smallest& 0.1265& 0.2401& 0.6334\\
      &Rest    & 0.1597& 0.2148& 0.6255\\
  \end{tabular}\end{center}
\end{table}

A natural question to ask is whether the above stylized facts are indeed related to 
each other and if so, is it possible for one to understand its origin.  
In the following, we will give an attempt to answer the first question
which would hopefully shed light on searching for an answer to the second question.  
This paper is organized as follows.
In section~\ref{sec:2}, we will give detailed analysis
of volatility clustering in financial time series.
In particular, we give arguments on what ingredient in financial time series 
is responsible for reproducing the nonlinear autocorrelations of price 
returns such as the one shown in figure~\ref{fig:nasdaq:af}.
We then introduce, in section~\ref{sec:3},
an index as a quantitative measure of volatility clustering in financial time series.
This would allow us to directly compare the degree of volatility clustering
across different financial time series.
The asymmetry between rises (gains) and falls (losses)
in the time series will be discussed in section~\ref{sec:4}. 
Section~\ref{sec:5} will be the summary and discussion.
In this work,
we have carried out the analysis on seven different representative financial time series.
They include (i) NASDAQ Composite Index (NASDAQ), (ii) Standard \& Poor's 500 index (S\&P500), 
(iii) Hang Seng Index (HSI), (iv) Microsoft stock price (MSFT),
(v) US Dollar/New Taiwan Dollar (USD/NTD),
(vi) Australian Dollar/New Taiwan Dollar (AUD/NTD)
and (vii) West Texas Intermediate (WTI). 
While we use NASDAQ as an example throughout the paper,
we will include the results of other financial time series in the appendix.  
	
\section{Volatility clustering and autocorrelation functions}
\label{sec:2}

\begin{figure}
  \centering{\includegraphics[width=0.8\textwidth]{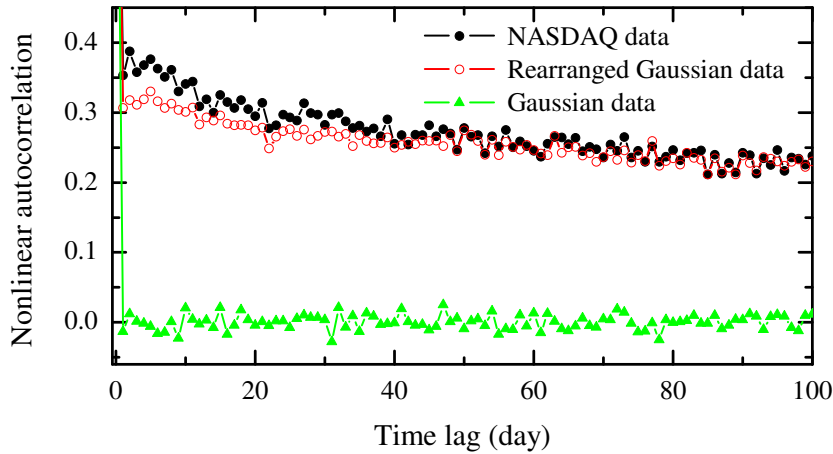}}
  \caption{\label{fig:simulation:af}
  The nonlinear autocorrelation function of the empirical data and
  of the rearranged Gaussian data.}
\end{figure}

We now begin our study by looking into the question of whether there
is a relationship among the heavy tails of return distributions,
volatility clustering and autocorrelation functions,
if the answer is yes, how they are related. 
Let us begin by asking the following question:
Is it necessary for one to have a heavy-tailed distribution
in order for the nonlinear autocorrelation function to exhibit the slow decay?
To answer this question, let us now assume that the return distribution
follow a Gaussian distribution instead of the empirical distribution shown
in figure~\ref{fig:nasdaq:pdf}.  
In this case, we assume the Gaussian distribution to have its mean and
standard deviation to be the same as the mean and the standard deviation
of the daily returns series in figure~\ref{fig:nasdaq:return}.
One can easily perform a simulation on this.
We now draw an equal number of returns from this Gaussian distribution
and call it the simulated data set.
After this is done, we sort both the empirical set and the simulated set
in the descending order of absolute returns.
We then substitute the values in the empirical data set by
the simulated data set one by one from the largest fluctuation to the smallest one
and calculate the nonlinear autocorrelation function of this rearranged Gaussian data.
The result is presented in figure~\ref{fig:simulation:af}.
For comparison, we also include the nonlinear autocorrelation of the empirical data
and the result from a pure Gaussian noise which is drawn from a Gaussian distribution
but without arranging the data according to the positions of empirical data set
like we do for the rearranged Gaussian data.
The pure Gaussian noise shows no temporal correlations as expected.
What is surprising is that the rearranged Gaussian returns
shows the same kind of slow decay behaviour as the empirical data set. 
On the other hand, if we randomize the temporal positions of the empirical returns,
namely, we reshuffle the original financial time series,
the result we obtain is always similar to the case of the pure Gaussian noise,
which means that there is no temporal correlation.
The above analysis therefore strongly suggests that
the heavy tails in return distributions are not responsible for 
the slow decay behaviour of the nonlinear autocorrelation functions.  

If the heavy tails in the distributions are not responsible for slow decay
in nonlinear autocorrelation functions,
what possible ingredients in the financial time series
would be responsible for such a slow decay behaviour.
We here try to provide an answer to this question.
Let us begin by looking at the clustering
of large price fluctuations in figure~\ref{fig:nasdaq:return}.
We begin by picking out the largest $p$\% (where $p$ is a constant)
fluctuations (whether they are positive or negative)
in the time series\footnote{
A similar treatment, the return interval approach~\cite{ref:RI1,ref:RI2,ref:RI3,ref:RI4},
is to pick the large fluctuations that are outside
$q$ standard deviations of the average value of the returns,
where $q$ is a pure number.}
and see whether their clustering behaviour would affect
the nonlinear autocorrelation function of the returns.
Since we are only interested in the clustering behaviour,
which in turn means the temporal positions but not the values of the 
large fluctuations in the financial time series,
we can here simply use 1 to represent the largest $p$\% fluctuations and 0
for all the other smaller fluctuations.
In this way, we will have a sequence which contains only 0 and 1.
This will in turn make our analysis much easier to interpret.
\begin{figure}
  \centering{\includegraphics[width=0.8\textwidth]{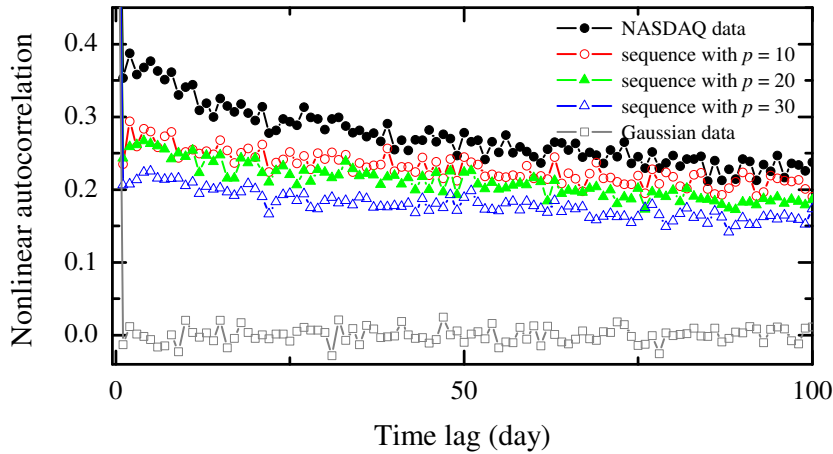}}
  \caption{\label{fig:seq:af}
  The nonlinear autocorrelation functions of the empirical data, the Gaussian data and
  the sequences of 1s and 0s with $p=10$, 20 and 30.
  $p$ here refers to the largest $p$\% fluctuations in the empirical returns
  and are represented by 1s while the rest are represented by 0s.}
\end{figure}
Figure~\ref{fig:seq:af} shows the nonlinear autocorrelation
of figure~\ref{fig:nasdaq:return}
using 1 for the largest $p$\% fluctuations and 0 for the rest.
We here include the results for sequences with $p=10$, 20 and 30.
To facilitate our discussion, we also include both the nonlinear autocorrelation
for empirical data and the Gaussian noise for comparison.
One can see that all these sequences show similar slow decay behaviour
as the original empirical data set, though with smaller values.  
This analysis thus shows that the positions of the large fluctuations 
are essential for a slow-decaying nonlinear autocorrelation function.  
Therefore, one can conclude that it is the clustering of large fluctuations
rather than the heavy tail in the return distribution 
which should be responsible for the slow decay behaviour of nonlinear autocorrelation functions.
This fact has also been observed in the other financial time series
in our study and the results are presented in the appendix.  
\begin{figure}
  \centering{\includegraphics[width=0.8\textwidth]{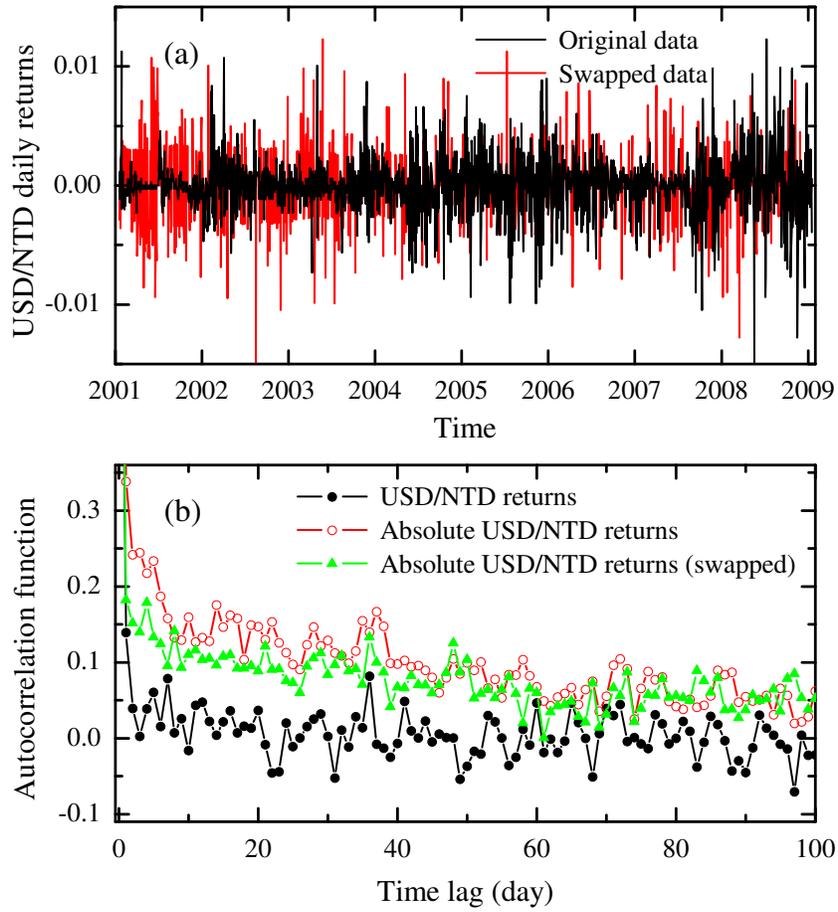}}
  \caption{\label{fig:USDNTD:af}
  (a) The historical daily return series for the currency exchange rate USD/NTD (black)
  and the series with the largest 20\% and smallest 20\% of the returns being swapped (red).
  (b) The nonlinear autocorrelations of the original returns (line with open circles)
  and the swapped returns (line with triangles).}
\end{figure}

Before we end this section, we would also like to make a further
study of the clustering of fluctuations in financial time series.
Instead of looking at the clustering of large fluctuations,
we now focus on the clustering of the small fluctuations in time series.
Since small fluctuations are smaller in value and
basically do not contribute to the nonlinear autocorrelation functions,
they are often left out in the discussion in the literature.
However, whether their temporal positions in a time series can have similar effects
as the large fluctuations is an interesting question that one can ask.
In figure~\ref{fig:USDNTD:af} (a), we plot the historical daily return time series
of the currency exchange rate USD/NTD from July 2, 2001 through June 30, 2009,
where the black line denotes the original empirical returns
while the red one represents the same set but with the largest 20\%
and smallest 20\% of the returns being swapped
\footnote{Our swapping procedure is described as follows.
We first swap the largest fluctuation (whether they are positive or negative)
with the smallest one in the time series
and then the second largest fluctuation with the second smallest one, and so on,
until the required percentage is achieved.}.
The nonlinear autocorrelations of the original empirical returns (line with open circles)
and the swapped returns (line with triangles) are drawn in figure~\ref{fig:USDNTD:af} (b).
We also include in this figure the autocorrelation function of the original returns for comparison.
One can see that although the line with triangles has values smaller than the original data set,
both lines have similar slow decay behaviour.
This in turn means that the clustering of small fluctuations in this returns series
has basically the same kind of feature as that of their large fluctuation counterparts.  
On the other hand, as we swap the large and small fluctuations in the other
six financial time series that we have been investigating,
the nonlinear autocorrelation functions of the swapped returns series
show no sign of slow decay.
They basically drop very fast,
similar to the kind of Gaussian noise in figure~\ref{fig:simulation:af}.
This interesting fact will be discussed in more detail in the next section 
as we introduce a clustering index to quantitatively study the 
clustering behaviour of different financial time series.
The introduction of this index would then allow us to directly compare
the degree of clustering across different financial time series.
\section{Quantitative measurement of volatility clustering}
\label{sec:3}
As mentioned above, in order to discuss the volatility clustering in a 
more quantitative way, it is better to introduce some parameters to 
quantitatively measure the volatility clustering of different financial 
time series that we can make comparison with.
We here introduce an index to quantify the volatility clustering in
the financial time series.
We begin by introducing a moving window with a certain window size
to scan through a given time series.
As an example, one can pick a window with size of $n$
(where $n$ is fixed throughout the scanning process) trading days.
Similar to what we have done in the previous section,
we can count the total number of trading days
that are among the largest $p$\% fluctuations in returns within this window
as we scan through the time series.
As we will see, 
one can interpret this as the degree of volatility clustering of the largest $p$\%
fluctuations with respect to this particular window with size $n$.  

\begin{figure}
  \centering{\includegraphics[width=0.8\textwidth]{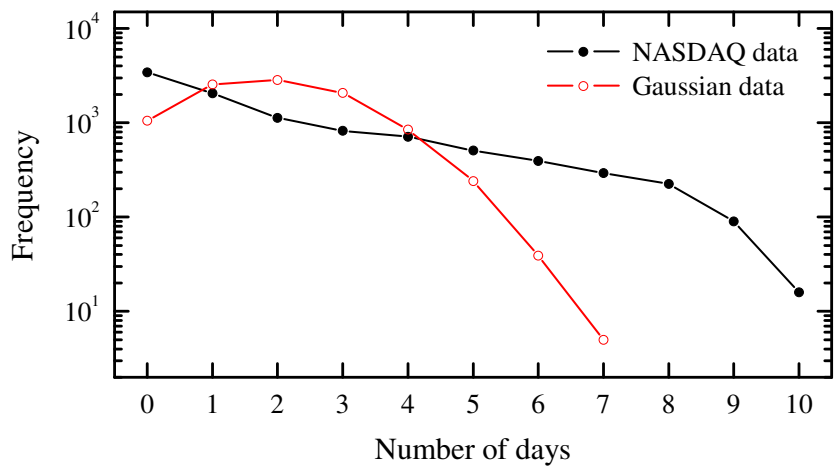}}
  \caption{\label{fig:clustering:distribution}
  The plot of the frequency distribution of the number of days with
  largest 20\% fluctuations within a window of 10 trading days.}
\end{figure}
Figure~\ref{fig:clustering:distribution}
is an illustration of the clustering of the largest 20\% fluctuations 
in figure~\ref{fig:nasdaq:return} with a window size of 10 trading days,
a span of two weeks in real daily life.
The statistics here is obtained by using the so called moving window method.
This means that we begin by putting the window on the first day
of the whole series and count the number of days among largest 20\% fluctuations
within this 10-day window.  
This is the first step.
We then move on to the second day of the whole series and again count the number
of days among largest 20\% fluctuations within this next 10-day window, the second step.  
We repeat the same procedure until we finish scanning through the whole time series.
The curve with full circles in figure~\ref{fig:clustering:distribution}
is a plot of the frequency distribution of the number of days
among the largest 20\% fluctuations within a 10-day period by using this moving window method.
To make it into a quantitative measure of the degree of clustering,
we need to compare it with a randomly generated time series for example, a Gaussian noise series.
The curve with open circles in figure~\ref{fig:clustering:distribution}
is the frequency distribution of the number of days
of the largest 20\% fluctuations within a 10-day period
from a simulated Gaussian noise series.
From figure~\ref{fig:clustering:distribution},
one can already visually tell the difference between these two curves.  
To be more concise, we take the ratio of the standard deviation of the 
number of days of the largest $p$\% fluctuations within the $n$-day window 
between the empirical and the simulated data sets.
Mathematically, it is defined as $R_n \equiv \sigma_e/\sigma_{\rm G}$,
where $\sigma_e$ and $\sigma_{\rm G}$ are the standard deviation of the
number of days of the largest $p$\% fluctuations within an $n$-day period
for the empirical and simulated Gaussian data sets respectively.
The larger the ratio is, the larger the degree of clustering will be.
This result can be understood easily.
The average number of days of largest $p$\% fluctuations within
a window size of $n$ is equal to $p \times n/100$.
This is true irrespective of whether it is the empirical data set or the simulated one.
One can indeed see this for the simulated data set which has a peak near this value.
However, if the time series displays the phenomenon of 
clustering of large fluctuations,
there will be a higher frequency of occurrence that the number of days
of the largest $p$\% fluctuations within this window
is much larger than the average value $p \times n/100$.
Similarly, there will also be a higher frequency of occurrence
that the number of days of the largest $p$\% fluctuations within this window
is much smaller than the average value $p \times n/100$.
This scenario will indeed be reflected in the value of the standard deviation 
of the frequency distribution in figure~\ref{fig:clustering:distribution}.
Thus, one can simply take the ratio of the standard deviation
of the empirical and simulated data sets to get a quantitative measure
of the degree of  clustering of the largest $p$\% fluctuations
of the financial time series that one is interested in.  

The ratio or index $R_n$ that we introduce here can in fact be studied analytically.
It has both theoretical upper and lower bounds and the standard deviation of the 
simulated Gaussian noise can also be calculated analytically.  
Let us first derive the theoretical value of the standard deviation of the
simulated Gaussian noise.
Recall from above that the mean value of the 
average number of days of the largest $p$\% fluctuations within
a $n$-day window is equal to $p \times n/100$.
For a total of $n$ days, the probability that there are $m$ days
with fluctuations among the largest $p$\% fluctuations can be written as, 
\begin{eqnarray}
  \frac{n!}{m!(n-m)!} P^m (1-P)^{n-m}    \,\, ,
\end{eqnarray}
where $P$ denotes $p/100$.
We here convert the percentage into decimals for simplicity.
The standard deviation of the average number of days of the 
largest $p$\% fluctuations within a $n$-day window is therefore equal to
\begin{eqnarray}
  \sigma_{\rm G} = \left[\sum_{m=0}^n (m - P n)^2 P^m (1-P)^{n-m}\right]^{1/2} = \sqrt{nP(1-P)} \,\, ,
\end{eqnarray}
which is the familiar result in statistics for the standard deviation of a sequence 
of $n$ random events with occurrence probability $P$.
The theoretical lower bound for the index corresponds to the case when the time series is
completely random, which is therefore equal to 1.  

To get a theoretical upper limit of the standard deviation of
the average number of days of the largest $p$\% fluctuations within a $n$-day 
period, we proceed as follows.
We look for the extreme case when all
the largest $p$\% fluctuations are ordered one after the other,
then followed by the rest of the data points
(one can of course reverse the order of the largest $p$\% fluctuations and the rest).
The first $p$\% of the data points will then be represented by 1
and the rest will be by 0, as what we have done in the above.
This is the case when we should have the largest possible degree of clustering.
If one plots this extreme case in figure~\ref{fig:clustering:distribution},
one will have two peaks in the frequency distribution function, one is at 0,
and the other is at $n$ (10 in the case in figure~\ref{fig:clustering:distribution}).  
Let us now use a window of size $n$ and begin with the first data point,
which is a 1, and count the $n$ data points in this window,
all of which are 1s (assuming that the length
of the time series $N$ is much longer than the window size $n$).  
Recall that we call this procedure to be step one.  
We then let the window slide to the next data point, the second step, and so on.  
As the moving window continues to move along the time series,
it will have moved $PN-n+1$ steps before it reaches the first 0.
We again have $P$ here to be equal to $p/100$ for simplicity.
As it continues to move along the time series,
the number of 1s will decrease
while the number of 0s will increase until the window consists of all 0s.
There are then $(1-P)N-n+1$ steps which has all 0s within the moving window.  
For the whole time series, we have a total of $N-n+1$ steps so we have
to average over these steps.  
It is now easy to calculate the standard deviation in this extreme case,
which is the square root of the expression in Eq. (\ref{eq:sigma}).  
Recall that the average 1s within the moving window is $Pn$.
We then have
\begin{eqnarray}\label{eq:sigma}
  &\frac{1}{N-n+1} \{ (PN-n)(n-P n)^2 + \left[ (1-P)N-n \right] (Pn)^2 + \sum_{m=0}^n (m-Pn)^2 \}\nonumber&\\
  &=\frac{1}{N-n+1} \{ n^2(N-n-1)P(1-P) + \frac{n(n+1)(2n+1)}{6} - n^3 \left[ P^2+(1-P)^2 \right] \} \,.\;&
\end{eqnarray}
In the limit $PN$ and $(1-P)N >> n$, the right hand side of Eq. (\ref{eq:sigma}) reduces to $n^2 P(1-P)$.
Therefore, the theoretical limit of the standard deviation $\sigma_{\lim}$ as $N$ goes to infinity is
\begin{equation}
  \sigma_{\lim} = \sqrt{n^2 P(1-P)} \,\, .
\end{equation}
The theoretical upper limit of $R_n$ is then equal to 
\begin{equation}
  R_{n}^{\lim} = \frac{\sigma_{\lim}}{\sigma_{\rm G}} = \sqrt{n} \,\, .
\end{equation}

\begin{figure}
  \centering{\includegraphics[width=0.8\textwidth]{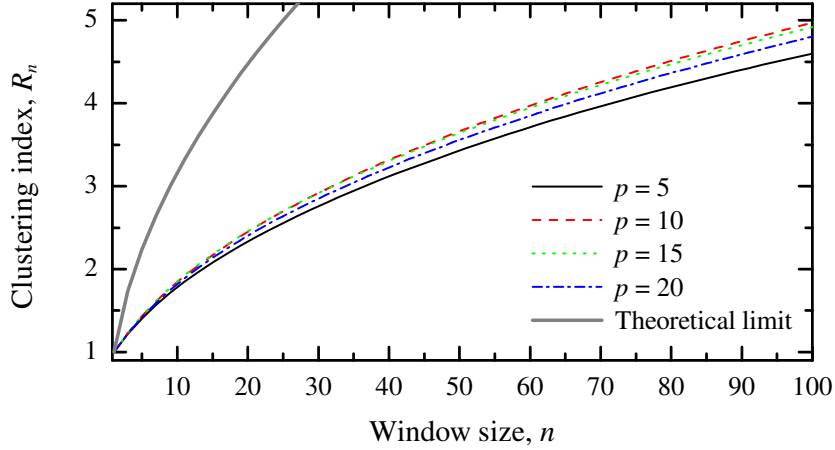}}
  \caption{\label{fig:clustering_index}
  The clustering index, $R_n$, for the NASDAQ return series with
  $p=5$ (solid), 10 (dash), 15 (dot) and 20 (dash dot).
  The theoretical limit of the index is drawn as a thick line for comparison.}
\end{figure}

Figure~\ref{fig:clustering_index} shows the value of the clustering index
for NASDAQ time series in figure~\ref{fig:nasdaq:return}
for various largest $p$\% fluctuations as a function of window size $n$.
The different curves represent the different largest $p$\% of the fluctuations in the time series.  
We have included here the results for $p=5$, 10, 15 and 20. 
Also included is the curve of the theoretical limit of the index.
The index values all start from unity when the window size $n$ corresponds to 1 trading day,
and gradually increase as the window size increases.  

\begin{figure}
  \centering{\includegraphics[width=0.8\textwidth]{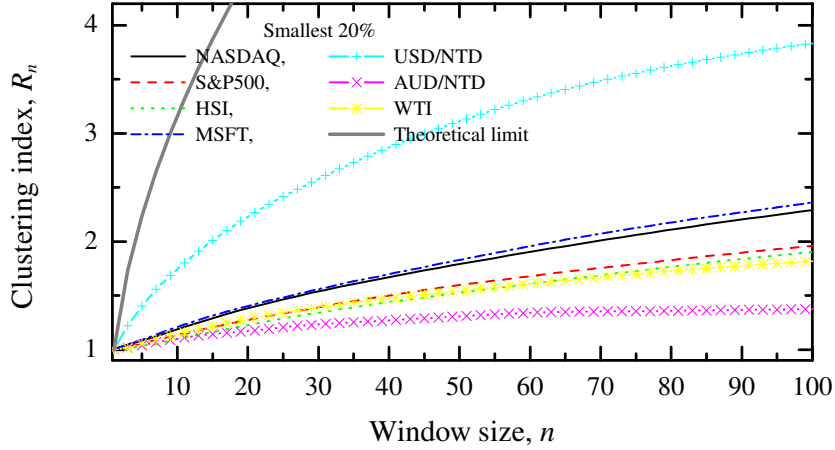}}
  \caption{\label{fig:bottom20}
    The clustering index for smallest 20\% returns of the
    NASDAQ, S\&P500, HSI, MSFT, USD/NTD, AUD/NTD and WTI series.
    The theoretical limit of the index is drawn as a thick line for comparison.}
\end{figure}

With the clustering index in hand, one can practically study the behaviour
of clustering of any sort of fluctuations in a financial time series.
Other than the largest $p$\% that we have looked into,
one can also look at the degree of clustering for small fluctuations.
To give the reader an idea of how one can use the index to study the properties
of financial time series,
we go back to a case which we considered in previous section.
Recall that we have studied a time series in which we swapped the largest $p$\%
and smallest $p$\% of the returns in the series,
as indicated in figure~\ref{fig:USDNTD:af}.
It turns out that the nonlinear autocorrelation function of the swapped data set
still exhibits similar slow decay behaviour.
On the other hand, we have analyzed the other six time series
that we consider in this paper and there is practically no such kind of slow
decay behaviour of the swapped data sets.
Using the index that we introduce here, the difference becomes clear.
Figure~\ref{fig:bottom20} shows the curves for the index value of the smallest 20\% returns
vs. window size in all the seven financial time series that we study in this work.
One can now see that the value of the index is rather small for each of the other 
six financial time series when compared with the curve for USD/NTD.
This means that the clustering of the smallest 20\% returns of these other financial time series 
indeed behave not much different from random sequences.
On the other hand, the clustering of the time series USD/NTD as shown 
in figure~\ref{fig:USDNTD:af} is significantly larger
which in turn reflects the slow decay behaviour
of the swapped data set in figure~\ref{fig:USDNTD:af}.
A possible explanation for the anomaly of clustering behaviour shown in the USD/NTD time series
might be a result from the stronger interference of Taiwan Central Bank than other countries.
Since Taiwan is an export-oriented country, the government
will attempt to regulate more frequently the fluctuations in the daily closing value of USD/NTD.
Whether this is the main reason for such an anomaly is not clear to us for the moment.
The above example thus suggests that the index that we introduce here is a good indicator
to quantify the degree of clustering of fluctuations in financial time series.
In the next section, we will see that the index that we introduce here indeed contains 
more information than people have previous observed in financial time series.
\section{Rise/Fall asymmetry}
\label{sec:4}
There exists discussions in the literature~\cite{ref:as1,ref:as2,ref:as3,ref:as4}
about the asymmetry of asset returns 
such as the skewness of the returns distribution in figure~\ref{fig:nasdaq:pdf}.
In the study of financial time series, one can for example,
ask whether there are more days that the returns are gains (rises) rather than losses (falls)
as some kind of asymmetry.
One can further ask whether the returns in gains would like to cluster more
or the other way round, how large the difference is,
and whether large fluctuations tend to cluster more, etc.
These can all be viewed as asymmetries in a financial time series.   
With the index introduced in previous section,
one should hopefully be able to extract more information
on the asymmetries in financial time series
and to study these asymmetries on a more quantitative basis.
To study the asymmetries in financial time series, let us first give the definitions here.
In the case of the asymmetry between the largest/smallest returns,
we adopt the notation that the largest and smallest fluctuations
refer to the absolute returns as before.
We first obtain the clustering index for the largest and smallest $p$\% returns.
The asymmetry of largest/smallest returns $A_{\rm{ls}}$ is then defined as
\begin{eqnarray}
  A_{\rm{ls}} = \frac{R_{\rm l} - R_{\rm s}}{R_{\rm l} + R_{\rm s}} \,\, ,
\end{eqnarray}
where $R_{\rm l}$ and $R_{\rm s}$ are the indices for the largest and smallest $p$\% 
fluctuations respectively.
This asymmetry will give us an idea whether the large fluctuations
or the small fluctuations would like to cluster more
as we increase the size of the moving window.
From this definition, it is clear that $A_{\rm{ls}}$ is equal to zero
when the window size is equal to 1, since there are
an equal number of largest and smallest fluctuations.

\begin{figure}
  \centering{\includegraphics[width=0.8\textwidth]{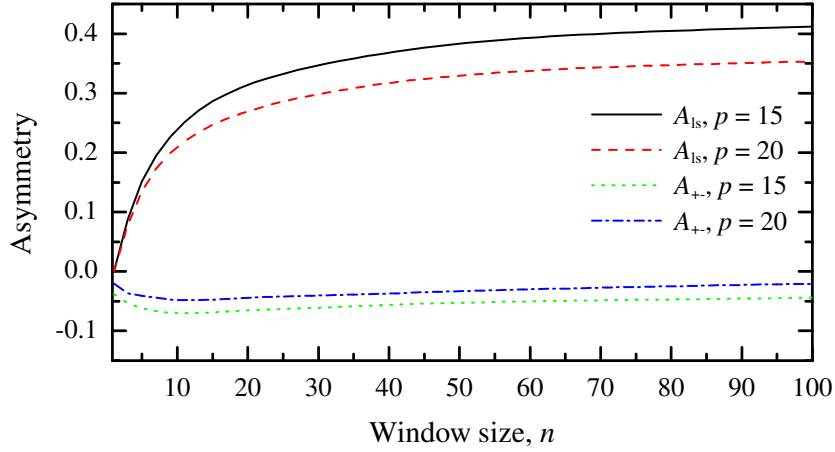}}
  \caption{\label{fig:nasdaq:asymmetry}
  The asymmetry $A_{\rm{ls}}$ and $A_{+-}$ with $p=15$ and 20 for NASDAQ return series.}
\end{figure}

In a similar fashion,
one can define the asymmetry between the largest positive and negative returns,
which we call $A_{+-}$ as follows
\begin{eqnarray}
  A_{+-} = \frac{R_+ - R_-}{R_+ + R_-} \,\, ,
\end{eqnarray}
where $R_+$ and $R_-$ are the indices for the largest positive and negative returns
respectively.
We should remind our reader here that in the case of $A_{+-}$,
we first pick up the largest $p$\% fluctuations from the absolute returns
and then separate the fluctuations (returns) into positive and negative categories.
In this way, we can see the asymmetry between the large positive and negative
returns as well as their degree of clustering.  
Notice that the asymmetries as defined above are bounded by 1 and -1.
Figure~\ref{fig:nasdaq:asymmetry} contains the plots of the asymmetries $A_{\rm{ls}}$ and $A_{+-}$
for $p$ = 15 and 20 for the NASDAQ time series in figure~\ref{fig:nasdaq:return}.
From the figure, it is easy to observe that the two curves for $A_{\rm{ls}}$ are always
positive, which means that the degree of clustering is more obvious
for large fluctuations than for small fluctuations in the NASDAQ times series.
On the other hand, the two curves for $A_{+-}$ are always below zero.
This reflects the fact that negative returns,
or big losses are likely to cluster together than big gains in the case of NASDAQ.
This is in agreement with some observations~\cite{ref:sf2,ref:as3}
indicating that there are more big losses rather than big gains in financial markets
since we have more big losses and these big losses are more likely to lump together.
We should remark here that the window size equals to 1
corresponds to the asymmetry of distribution of returns in figure~\ref{fig:nasdaq:pdf}.
In the case of NASDAQ, the asymmetry is negative.
There are however,
examples of financial time series that the asymmetry for the probability 
density function is positive and they are included in the appendix below.
By increasing the size of the moving window, one can also study the asymmetry of 
returns with respect to the clustering of large and small fluctuations.
Therefore, the use of the index to study asymmetries in financial time series
allows one to extract more information comparing to conventional methods.  

The asymmetry between rises (gains) and falls (losses) in returns can also be observed
in the probability of the occurrence of large and small fluctuations following
the occurrence of large or small ones on the previous day,
which is shown in table~\ref{tab:probability:as}.
Unlike table~\ref{tab:probability},
we now separate the rise and fall of the fluctuations into separate categories.
Each row in this table sums to unity as before.
In this way, one can easily detect the difference.
\begin{table}
  \caption{\label{tab:probability:as}
  The probability of the occurrence of large and small rise/fall following
  the occurrence of large or small rise/fall on the previous day (the first column).}
  \begin{center}\begin{tabular}{@{}ll|lll}
      \hline
      &20\%&Largest (rise/fall)&Smallest (rise/fall)&Rest (rise/fall)\\
      \hline
      &Largest (rise) & 0.2054/0.1514& 0.0551/0.0724& 0.3319/0.1838\\
      &Largest (fall) & 0.1856/0.2438& 0.0365/0.0681& 0.2122/0.2538\\
      &Smallest (rise)& 0.0451/0.0573& 0.1437/0.1023&	0.3624/0.2892\\
      &Smallest (fall)& 0.0790/0.0767& 0.1226/0.1100&	0.3471/0.2646\\
      &Rest (rise)    & 0.0742/0.0582& 0.1269/0.0934&	0.4475/0.1998\\
      &Rest (fall)    & 0.0737/0.1255& 0.1183/0.0888&	0.2888/0.3049\\
  \end{tabular}\end{center}
\end{table}

\section{Summary and discussion}
\label{sec:5}
In this paper, we have made a detailed analysis of the stylized facts in financial time series. 
We have found that the slow decay behaviour is directly related to the degree of clustering
of the large fluctuations (absolute returns) within the financial time series
while the heavy tails in return distributions do not seem to play a role here.
We have also introduced an index to quantitatively measure the 
clustering behaviour of fluctuations in financial time series
and have given examples to demonstrate its advantages over the conventional methods.
This index has both theoretical lower and upper bounds.
It is equal to unity if the fluctuations are independent,
identically distributed within the financial time series.
On the other hand, its upper bound can also be analytically calculated
and in the limit when the time scale of the given series is much longer
than the window size $n$, the index $R_n$ is simply equal to $\sqrt{n}$. 
With this index in hand,
one not only can study the asymmetry of the asset returns but also the effect of 
clustering on the asymmetry properties in financial time series.
One can see that the larger fluctuations tend to cluster more than the smaller ones.
Similarly, big losses tend to lump together more severely than big gains.
These findings should be helpful to people who make investments in financial markets.
Indeed, the clustering index introduced might also be employed to investigate the
clustering behaviour of other nonlinear systems.  

Whether the clustering of large fluctuations in financial time series is from
long time memory effect or other effects such as human psychology is not clear 
to us at the moment.
Indeed, if only the clustering of large fluctuations is concerned,
one is able to find complex systems that are random in nature
but can possess the property of large degree of clustering.
It is also possible to find simple ways which can give both large degrees of clustering
and slow-decaying nonlinear autocorrelations in a simulated time series.
We will discuss these issues in a future publication~\cite{ref:LM}.  
\section*{Acknowledgments}
This work was supported in part by the National Science Council of Taiwan
under grants NSC\#98-2120-M-001-002 and NSC\#97-2112-M-001-008-MY3.
\appendix
\section*{Appendix}\label{app:appendix}
\setcounter{section}{1}
\setcounter{figure}{0}
\setcounter{table}{0}
For comparison,
we here show the results of other financial time series as mentioned in section~\ref{sec:1}
with the same procedures introduced in this paper.
They include (i) Standard \& Poor's 500 index (S\&P500, from January 4, 1950 through June 30, 2009), 
(ii) Hang Seng Index (HSI, from January 2, 1987 through June 30, 2009),
(iii) Microsoft stock price (MSFT, from March 4, 1986 through June 30, 2009),
(iv) US Dollar/New Taiwan Dollar (USD/NTD, from July 2, 2001 through June 30, 2009),
(v) Australian Dollar/New Taiwan Dollar (AUD/NTD, from July 2, 2001 through June 30, 2009)
and (vi) West Texas Intermediate (WTI, from January 6, 1986 through June 30, 2009). 

\begin{figure}[HT]
  \centering{\includegraphics[width=0.96\textwidth]{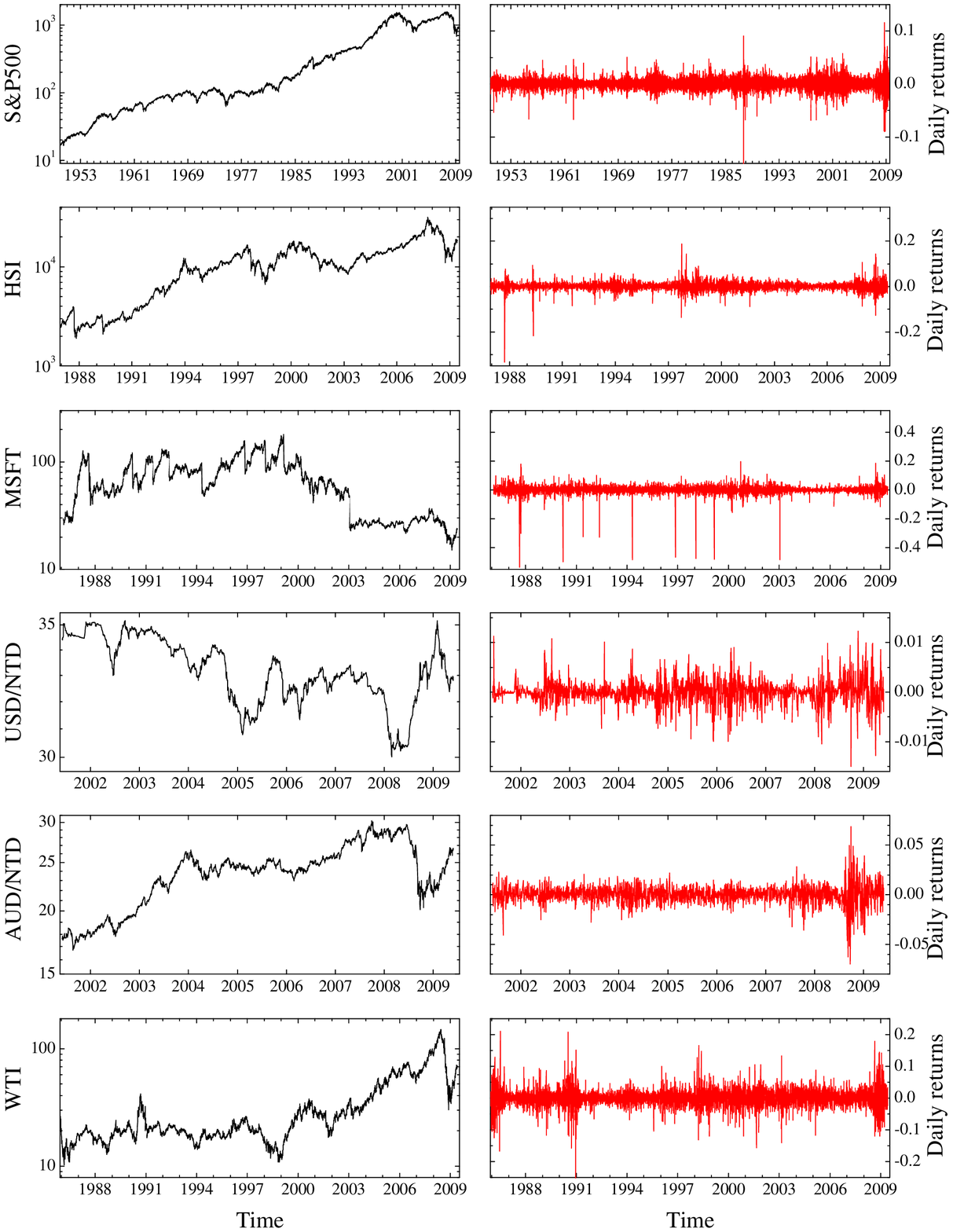}}
  \caption{\label{fig:all:return}
  The left panels are plots of the historical daily closing value for S\&P500, HSI, MSFT,
  USD/NTD, AUD/NTD and WTI (from top to bottom),
  while the right panels show the daily returns for each of these time series.}
\end{figure}
\begin{figure}
  \centering{\includegraphics[width=0.96\textwidth]{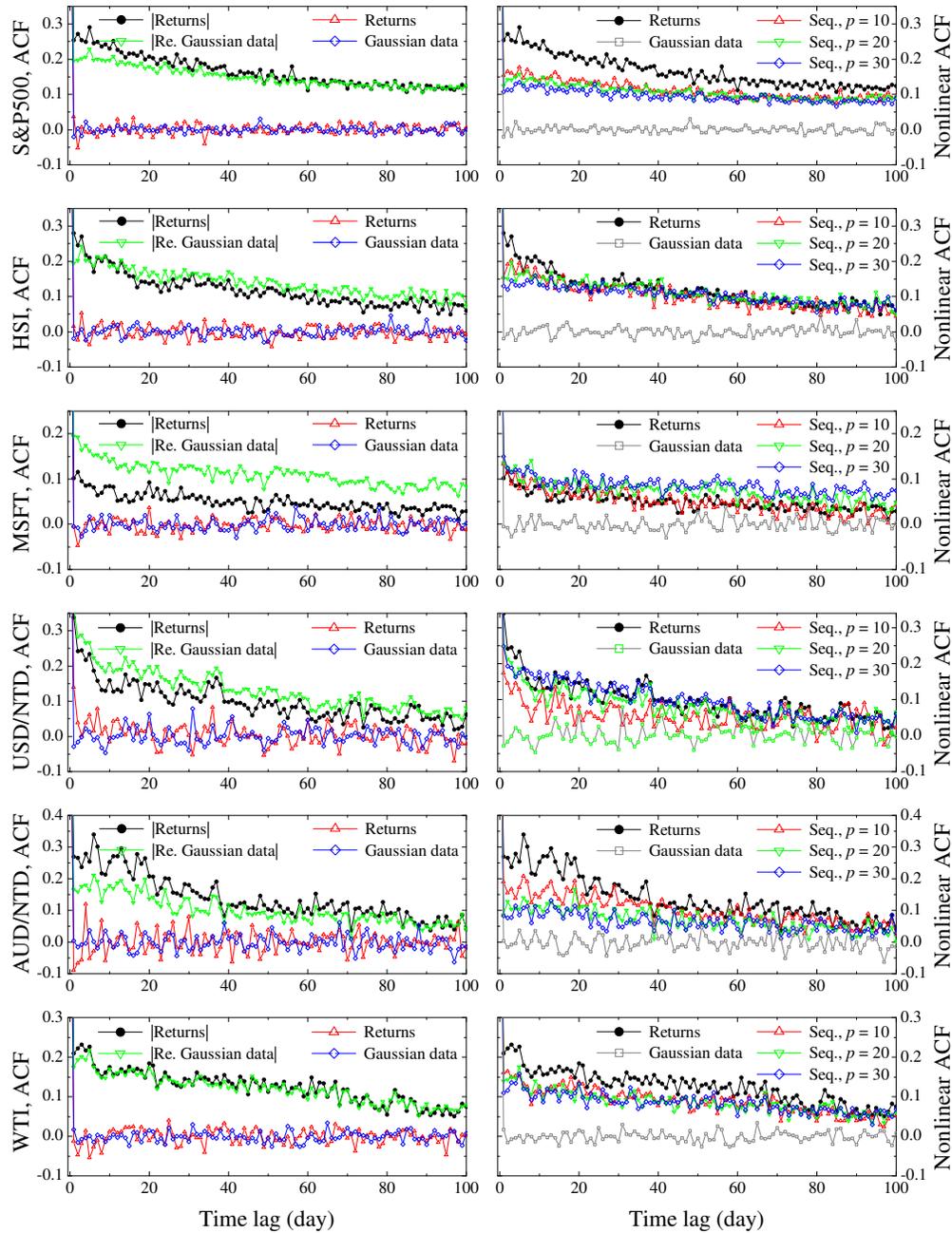}}
  \caption{\label{fig:all:AF}
  The left panels are plots of the autocorrelation functions
  for S\&P500, HSI, MSFT, USD/NTD, AUD/NTD and WTI returns (from top to bottom),
  while the right panels show the nonlinear autocorrelations of the sequences of 
  1s and 0s with $p=10$, 20 and 30 as extracted from these returns.}
\end{figure}
\begin{table}[HT]
  \caption{\label{tab:all:probability}
  The probability of the occurrence of large and small fluctuations following
  the occurrence of large or small ones on the previous day (the first column).}
  \begin{center}\begin{tabular}{@{}ll|lll}
      \hline
      &S\&P500, 20\% &Largest& Smallest& Rest\\
      \hline
      &Largest & 0.2999&	0.1556&	0.5445\\
      &Smallest& 0.1440&	0.2305&	0.6255\\
      &Rest    & 0.1853&	0.2046&	0.6101\\
\\
      \hline
      &HSI, 20\% &Largest& Smallest& Rest\\
      \hline
      &Largest & 0.3226&	0.1568&	0.5206\\
      &Smallest& 0.1587&	0.2126&	0.6287\\
      &Rest    & 0.1730&	0.2101&	0.6169\\
\\
      \hline
      &MSFT, 20\% &Largest& Smallest& Rest\\
      \hline
      &Largest & 0.3087&	0.1327&	0.5586\\
      &Smallest& 0.1354&	0.2419&	0.6227\\
      &Rest    & 0.1852&	0.2085&	0.6063\\
\\
      \hline
      &USD/NTD, 20\% &Largest& Smallest& Rest\\
      \hline
      &Largest & 0.3985&	0.0777&	0.5238\\
      &Smallest& 0.0501&	0.4010&	0.5489\\
      &Rest    & 0.1840&	0.1724&	0.6436\\
\\
      \hline
      &AUD/NTD, 20\% &Largest& Smallest& Rest\\
      \hline
      &Largest & 0.2700&	0.1600&	0.5700\\
      &Smallest& 0.1825&	0.2200&	0.5975\\
      &Rest    & 0.1827&	0.2068&	0.6105\\
\\
      \hline
      &WTI, 20\% &Largest& Smallest& Rest\\
      \hline
      &Largest & 0.3120&	0.1762&	0.5118\\
      &Smallest& 0.1603&	0.2025&	0.6372\\
      &Rest    & 0.1761&	0.2071&	0.6168\\
  \end{tabular}\end{center}
\end{table}
\begin{figure}
  \centering{\includegraphics[width=0.96\textwidth]{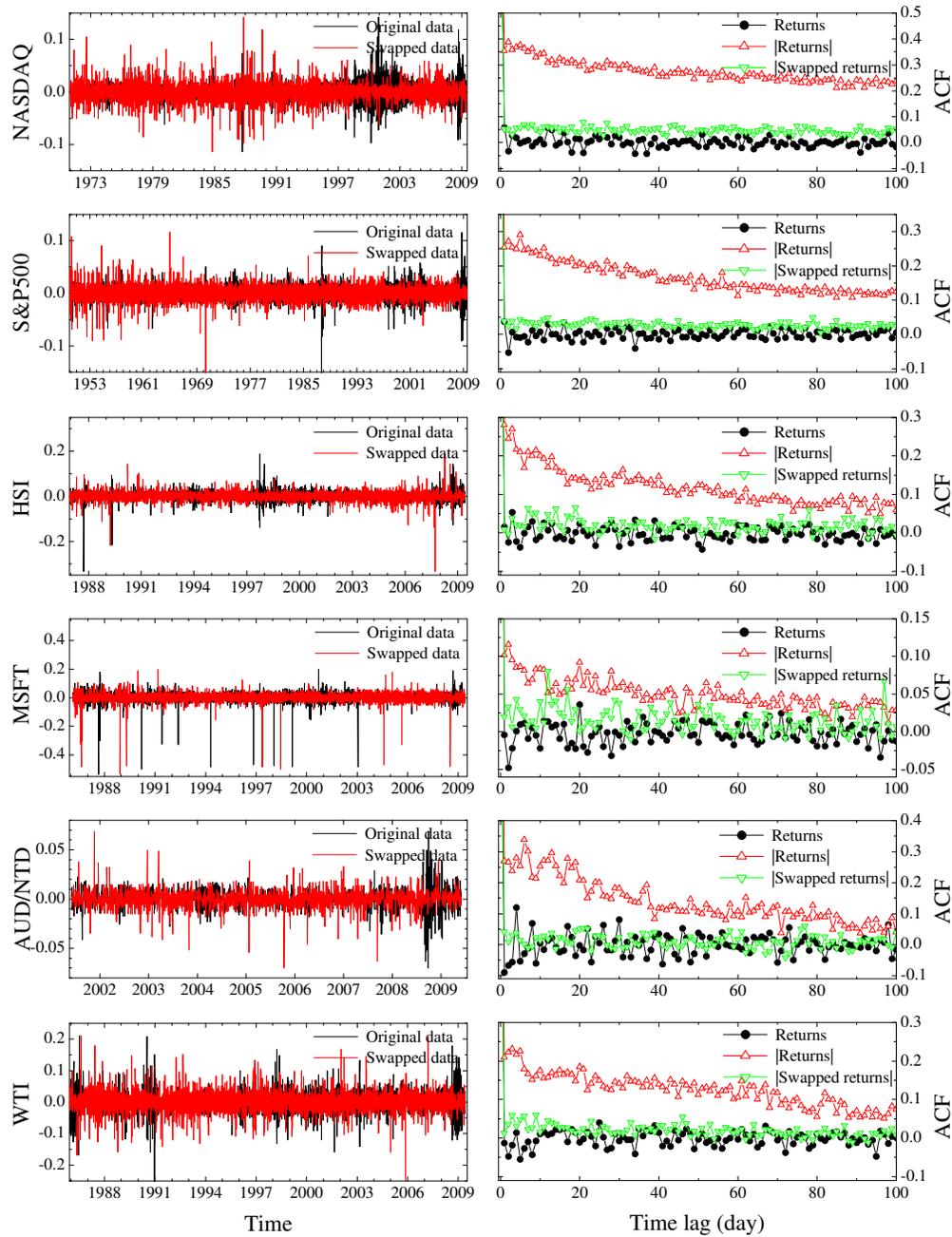}}
  \caption{\label{fig:all:swap}
  The left panels show the original returns series
  and the $20\%$ swapped series
  for NASDAQ, S\&P500, HSI, MSFT, AUD/NTD and WTI (from top to bottom),
  while the right panels plot the autocorrelations
  for each of these time series.}
\end{figure}
\begin{figure}
  \centering{\includegraphics[width=0.96\textwidth]{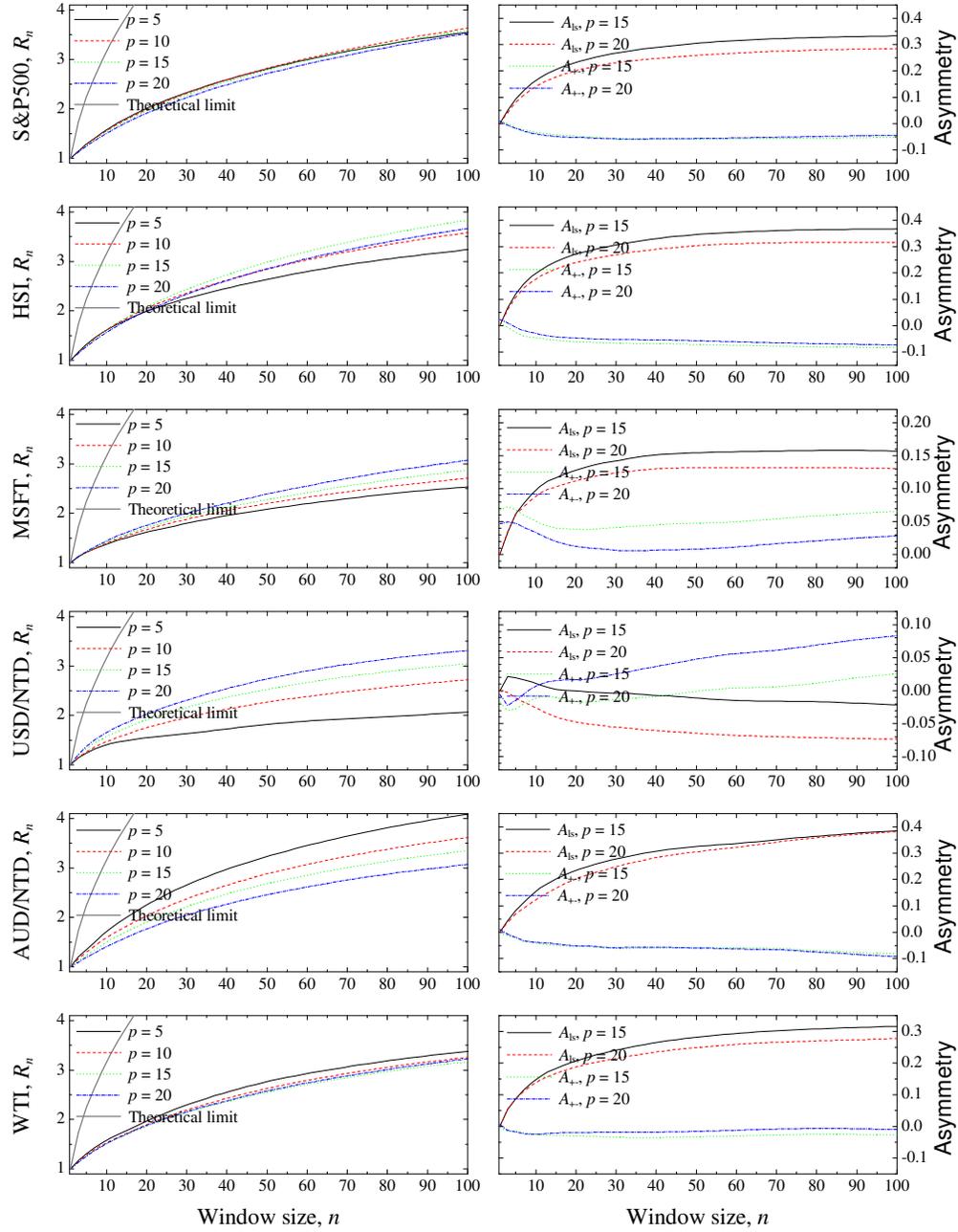}}
  \caption{\label{fig:all:CI}
  The left panels show the clustering index $R_n$ for S\&P500, HSI, MSFT,
  USD/NTD, AUD/NTD and WTI (from top to bottom),
  while the right panels are the asymmetries $A_{\rm{ls}}$ and $A_{+-}$ for each of 
  these time series.}
\end{figure}
\begin{table}
  \caption{\label{tab:probability:asall_1}
  The probability of the occurrence of large and small rise/fall following
  the occurrence of large or small rise/fall on the previous day (the first column).}
  \begin{center}\begin{tabular}{@{}ll|lll}
      \hline
      &S\&P500, 20\%&Largest (rise/fall)&Smallest (rise/fall)&Rest (rise/fall)\\
      \hline
      &Largest (rise) & 0.1634/0.0954&	0.0922/0.0797&	0.3150/0.2543\\
      &Largest (fall) & 0.1598/0.1831&	0.0697/0.0690&	0.2602/0.2582\\
      &Smallest (rise)& 0.0678/0.0622&	0.1307/0.1195&	0.3497/0.2701\\
      &Smallest (fall)& 0.0808/0.0794&	0.1053/0.1025&	0.3210/0.3110\\
      &Rest (rise)    & 0.0963/0.0764&	0.1224/0.0979&	0.3590/0.2480\\
      &Rest (fall)    & 0.0864/0.1139&	0.1003/0.0857&	0.3047/0.3090\\
\\
      \hline
      &HSI, 20\%&Largest (rise/fall)&Smallest (rise/fall)&Rest (rise/fall)\\
      \hline
      &Largest (rise) & 0.1684/0.1088&	0.1003/0.0884&	0.2908/0.2433\\
      &Largest (fall) & 0.1799/0.1932&	0.0625/0.0587&	0.2973/0.2084\\
      &Smallest (rise)& 0.0904/0.0638&	0.1099/0.0993&	0.3351/0.3015\\
      &Smallest (fall)& 0.0888/0.0743&	0.1087/0.1069&	0.3243/0.2970\\
      &Rest (rise)    & 0.0912/0.0642&	0.0996/0.1210&	0.3388/0.2852\\
      &Rest (fall)    & 0.0842/0.1091&	0.1103/0.0886&	0.3055/0.3023\\
\\
      \hline
      &MSFT, 20\%&Largest (rise/fall)&Smallest (rise/fall)&Rest (rise/fall)\\
      \hline
      &Largest (rise) & 0.1726/0.1217&	0.0786/0.0586&	0.2865/0.2820\\
      &Largest (fall) & 0.1879/0.1385&	0.0835/0.0436&	0.2960/0.2505\\
      &Smallest (rise)& 0.0767/0.0737&	0.1293/0.0917&	0.3038/0.3248\\
      &Smallest (fall)& 0.0629/0.0530&	0.1257/0.1434&	0.2888/0.3262\\
      &Rest (rise)    & 0.1045/0.0858&	0.1148/0.0886&	0.2983/0.3080\\
      &Rest (fall)    & 0.0963/0.0839&	0.1235/0.0901&	0.3082/0.2980\\
    \end{tabular}\end{center}
\end{table}
\begin{table}
  \caption{\label{tab:probability:asall_2}
  The probability of the occurrence of large and small rise/fall following
  the occurrence of large or small rise/fall on the previous day (the first column).}
  \begin{center}\begin{tabular}{@{}ll|lll}
      \hline
      &USD/NTD, 20\%&Largest (rise/fall)&Smallest (rise/fall)&Rest (rise/fall)\\
      \hline
      &Largest (rise) & 0.2183/0.1371&	0.0863/0.0051&	0.2741/0.2791\\
      &Largest (fall) & 0.1584/0.2822&	0.0594/0.0050&	0.2525/0.2425\\
      &Smallest (rise)& 0.0268/0.0368&	0.2542/0.1171&	0.2843/0.2808\\
      &Smallest (fall)& 0.0112/0.1899&	0.0894/0.1117&	0.1676/0.4302\\
      &Rest (rise)    & 0.0711/0.1438&	0.0321/0.4264&	0.2640/0.0626\\
      &Rest (fall)    & 0.0651/0.1127&	0.1320/0.0511&	0.2975/0.3416\\
\\
      \hline
      &AUD/NTD, 20\%&Largest (rise/fall)&Smallest (rise/fall)&Rest (rise/fall)\\
      \hline
      &Largest (rise) & 0.0780/0.1220&	0.1024/0.0976&	0.2585/0.3415\\
      &Largest (fall) & 0.2359/0.1077&	0.0667/0.0513&	0.3231/0.2153\\
      &Smallest (rise)& 0.0679/0.1176&	0.1222/0.0950&	0.3620/0.2353\\
      &Smallest (fall)& 0.1056/0.0722&	0.1500/0.0778&	0.3500/0.2444\\
      &Rest (rise)    & 0.0783/0.0934&	0.1130/0.0994&	0.3178/0.2981\\
      &Rest (fall)    & 0.1067/0.0899&	0.1086/0.0918&	0.3633/0.2397\\
\\
      \hline
      &WTI, 20\%&Largest (rise/fall)&Smallest (rise/fall)&Rest (rise/fall)\\
      \hline
      &Largest (rise) & 0.1331/0.1514&	0.0998/0.0815&	0.2396/0.2946\\
      &Largest (fall) & 0.1573/0.1829&	0.0940/0.0769&	0.2615/0.2274\\
      &Smallest (rise)& 0.0801/0.0740&	0.1193/0.0952&	0.3127/0.3187\\
      &Smallest (fall)& 0.0860/0.0822&	0.0994/0.0880&	0.3614/0.2830\\
      &Rest (rise)    & 0.0897/0.0756&	0.1178/0.1032&	0.3252/0.2885\\
      &Rest (fall)    & 0.0969/0.0910&	0.1163/0.0757&	0.3259/0.2942\\
    \end{tabular}\end{center}
\end{table}
\section*{References}
\bibliographystyle{elsarticle-num}
\bibliography{volatility_clustering}
\end{document}